%% file: Tcc_RQCD.tex
\documentclass[twocolumn,groupedaddress,aps,nofootinbib,prd,superscriptaddress,preprintnumbers]{revtex4-1} 
\usepackage{amsmath} 
\usepackage{amssymb}
\usepackage{epsfig}
\usepackage{color} 
\usepackage{multirow}
\usepackage{soul} 
\usepackage[utf8]{inputenc}

\begin{document}

\title{Signature of a doubly charm tetraquark pole in $DD^*$ scattering on the lattice}

 \author{M. Padmanath}
 \email{pmadanag@uni-mainz.de, papppan@gmail.com}
      \affiliation{ Helmholtz Institut Mainz,   Mainz, Germany}
         \affiliation{ GSI Helmholtzzentrum f\"ur Schwerionenforschung, Darmstadt, Germany }  
   \author{S. Prelovsek}
 \email{sasa.prelovsek@ijs.si}
   \affiliation{ Faculty of Mathematics and Physics, University of Ljubljana, Ljubljana, Slovenia }
    \affiliation{Jozef Stefan Institute, Ljubljana, Slovenia } 

\begin{abstract}  
The doubly charm tetraquark with flavor $cc\bar u\bar d$ and isospin $I\!=\!0$ is investigated 
by calculating the $DD^*$ scattering amplitude with lattice QCD. The simulation is done on 
CLS ensembles with dynamical $u/d,s$ quarks and $m_\pi\simeq 280~$MeV for two charm quark 
masses, one slightly larger and one slightly lower than the physical value. The scattering 
amplitudes for partial waves $l=0,1$ are extracted near-threshold via the L\"uscher's method 
by considering systems with total momenta $PL/(2\pi)=0,1,\sqrt{2},2$ on two spatial volumes. 
A virtual bound state pole in the $DD^*$ scattering amplitude with $l=0$ is found 
$9.9_{-7.1}^{+3.6}~$MeV below $DD^*$ threshold for the charm quark mass closer to the physical 
value. This pole is likely related to the doubly charm tetraquark discovered by LHCb less than 
$1~$MeV below $D^0D^{*+}$ threshold. Future lattice simulations closer to the continuum limit 
and physical quark masses would be valuable to establish this connection systematically. 
\end{abstract}

\maketitle
\preprint{MITP/22-018}

{\it Introduction}: The LHCb collaboration  recently discovered a  doubly charmed tetraquark 
$T_{cc}$ with flavor $cc\bar u\bar d$ just $0.36(4)~$MeV below $D^0D^{*+}$ threshold 
\cite{LHCb:2021vvq,LHCb:2021auc}\footnote{The mass obtained from the pole position in 
Ref. \cite{LHCb:2021auc} is quoted.}. Its flavor is based  on the decay channel 
$D^0D^0\pi^+$ and it has isospin $I=0$ since no state was found  in the  decay 
$D^0D^+\pi^+$. The total spin and parity $J^P$ have not  been determined from experiment.
This is the longest-lived hadron discovered with explicitly exotic quark content. 
It has striking similarities with the well-known $X(3872)$ \cite{Belle:2003nnu} that lies 
very close to $D^0\bar D^{*0}$ threshold. Here we aim at the theoretical investigation of 
near-threshold exotics from first principles.

Several phenomenological models predicted a doubly charm tetraquark $cc\bar u\bar d$ with 
$I=0$ and $J^P=1^+$ within an energy range $\pm 100~$MeV around the $DD^*$ threshold, 
{\it e.g.} \cite{Ader:1981db,Heller:1986bt,Carlson:1987hh,Navarra:2007yw,Ebert:2007rn,Karliner:2017qjm,
Eichten:2017ffp,Janc:2004qn,Carames:2011zz}. 
Many of these models have a possibility to identify a bound state but not a resonance. 
One of the more sophisticated quark model calculations predicted the bound state $1.6\pm 1.0~$MeV 
below $DD^*$ threshold and concluded that the molecular Fock component dominates over the 
diquark antidiquark component \cite{Janc:2004qn}. Within a molecular picture, a light vector 
meson exchange is argued to induce attraction \cite{Feijoo:2021ppq,Dong:2021bvy}, whereas 
one-pion exchange induces slight repulsion \cite{Du:2021zzh}. The binding energy of a bound 
state in the $QQ\bar u\bar d$ system is found to decrease with decreasing heavy quark mass $m_Q$ 
and with increasing light quark mass $m_{u,d}$ 
\cite{Junnarkar:2018twb,Pflaumer:2021ong,Francis:2016hui,Francis:2018jyb,Leskovec:2019ioa,
Frances:2022,Hudspith:2020tdf,Bicudo:2015vta,Karliner:2017qjm,Janc:2004qn,Francis:2021vrr}. 
Thus the doubly bottom tetraquarks $bb\bar u\bar d$ and $bb\bar u\bar s$ with $J^P\!=\!1^+$ are
deeply bound according to variety of theoretical approaches 
\cite{Junnarkar:2018twb,Pflaumer:2021ong,Francis:2016hui,Leskovec:2019ioa,
Frances:2022,Bicudo:2015vta,Karliner:2017qjm,Janc:2004qn}, whereas $cc\bar u\bar d$ is 
expected on the verge of binding and requires a careful theoretical study within QCD.  

In order to theoretically confirm the existence of a doubly charmed tetraquark from first-principles, 
one has to establish a pole in the corresponding scattering amplitude $t(E_{cm})$. This is 
particularly important in finite-volume formulations, such as lattice QCD, since this state does 
not lie well below the threshold but is expected near threshold. Lattice QCD represents the 
only non-perturbative first-principles approach with quantifiable systematic and statistical 
uncertainties to study QCD in the hadronic regime. It enables the determination of the scattering 
amplitudes from deviations of finite-volume energies from the non-interacting scenario 
\cite{Luscher:1990ux}. However, the scattering amplitude in this channel has not been determined 
using lattice simulations yet. The lattice study in Ref. \cite{Junnarkar:2018twb} extracted the 
finite-volume energy of the ground state using meson-meson and diquark-antidiquark interpolators 
for a wide range of $m_{\pi}\geq 260~$MeV and three lattice spacings. The continuum and chiral 
extrapolations lead to an energy level $-23\pm 11~$MeV relative to the $DD^*$ threshold. This 
indicates the presence of interactions between $D$ and $D^*$, but does not prove the existence 
of a pole. The finite-volume energies have been extracted in Ref. \cite{Cheung:2017tnt} and the 
ground state energy was found to be consistent with the $DD^*$ threshold. 

This letter investigates if a state with flavor $cc\bar u\bar d$, $I\!=\!0$, and $J^P\!=\!1^+$ 
exists in the vicinity of $DD^*$ threshold. For this purpose, $DD^*$ scattering amplitude 
$t(E_{cm})$ near threshold is extracted within lattice QCD for the first time. It  
is determined from finite-volume energies via the L\"uscher's method \cite{Luscher:1990ux}. 
The $D^*$ does not decay strongly to $D\pi$ at the simulated $m_\pi\!\simeq\! 280~$MeV and 
the analyzed energy region is below the $DD\pi$ and $D^*D^*$ thresholds, therefore we consider 
one-channel $DD^*$ scattering. We demonstrate that the scattering amplitude
indeed has a pole at $E_{cm}^p$ slightly below threshold.  
 \vspace{0.1cm}

First we present the calculation of the energy levels. Then we discuss the extraction of the 
scattering amplitude and the poles in it.
  \vspace{0.2cm}
  
{\it  Ensembles and single-hadron masses}: We utilize two ensembles with $u/d,s$ dynamical quarks 
provided by the Coordinated Lattice Simulations consortium \cite{Bruno:2014jqa,Bali:2016umi}. 
The lattice spacing is $a=0.08636(98)(40)$~fm, $m_u$ and $m_d$ are degenerate and heavier than 
in Nature, corresponding to   $m_\pi\!=\!280(3)~$MeV. There are 255 configurations on spatial 
volume $ N_L^3= 24^3$ and 492 configurations on $ 32^3$ \cite{Bruno:2016plf}. The scattering 
amplitude is extracted for two values of the charm quark mass, one slightly heavier than physical 
and one slightly lighter \cite{Piemonte:2019cbi}. The masses of the relevant hadrons $D$ and 
$D^*$ are presented in Table \ref{tab:results}. The heavier charm quark mass is closer to the 
physical value and provides our main result.    
   
\begin{table*}[tb]
  \begin{center}
    \begin{tabular}{c | ccc | c|c | c}
      \hline
      ID & $\vec P$ & $LG$ & $\Lambda^P$  & $J^P$ & $l$ & interpolators: $M_1(\vec p_1^{~2})M_2(\vec p_2^{~2})$ \\
      \hline
      1 & $(0,0,0)$ & $O_h$ & $T_1^+$ & $1^+$ & $0,2$  & $D(0)D^*(0),~ D(1)D^*(1)~[2],~D^*(0)D^*(0)$\\
      2 & $(0,0,0)$ & $O_h$ & $A_1^-$ & $0^-$ & $1$ & $ D(1)D^*(1)$\\
      3 & $(0,0,1)\tfrac{2\pi}{L}$ & $\mathrm{Dic}_4$ & $A_2$ & $0^-,1^+,2^-$ & $0,1,2$ & $D(0)D^*(1),~D(1)D^*(0)$  \\
      4 & $(1,1,0)\tfrac{2\pi}{L}$ & $\mathrm{Dic}_2$ & $A_2$ & $0^-,1^+,2^-,2^+$ & $0,1,2$  & $D(0)D^*(2),~D(1)D^*(1)~[2],~D(2)D^*(0)$  \\   
      5 & $(0,0,2)\tfrac{2\pi}{L}$ & $\mathrm{Dic}_4$ & $A_2$ & $0^-,1^+,2^-$ & $0,1,2$ & $D(1)D^*(1)$ \\
      \hline
  \end{tabular}
  \end{center}
  \caption{Total momenta $\vec P$, spatial lattice symmetry group ($LG$), irreducible representations 
($\Lambda^P$) and interpolators considered for the system $cc\bar u\bar d$, together with total 
spin-parity $J^P$ and partial-wave $l$ of $DD^*$ scattering that contributes to each irrep (only 
$J,l\leq 2$ are listed). The interpolators are denoted by $[2]$ when two linearly independent 
combinations of momenta and polarizations are employed, e.g. $O_{l=0,2}$ for $D(1)D^*(1)$ in $T_1^+$ \cite{Suppl}. }
  \label{tab:irreps}
\end{table*}
  \vspace{0.2cm}
 
\begin{figure}[tbh!]
\begin{center}
\includegraphics[height=7.5cm,width=8.3cm]{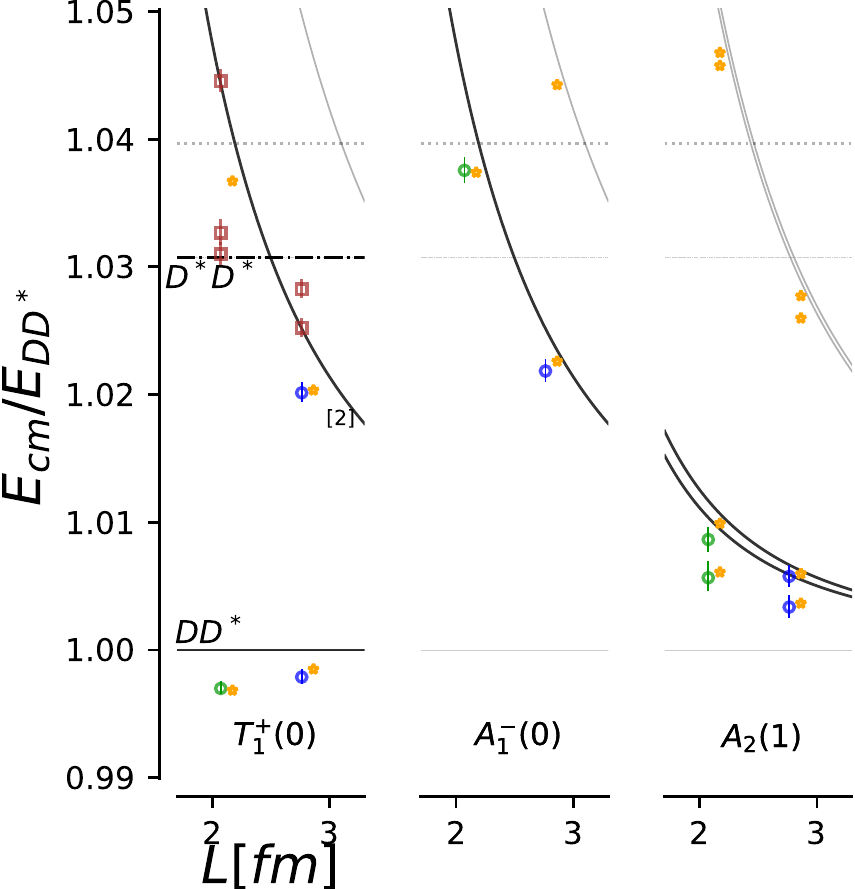}  
\caption{The center-of-momentum energy $E_{cm}=(E^2-\vec P^2)^{1/2}$ of the $cc\bar u\bar d$ system 
normalized by $E_{DD^*}\equiv m_D+m_{D^*}$, for the heavier charm quark mass in various finite-volume 
irreps. The lattice energy levels are shown by large circles and squares: the scattering analysis 
employs the blue and green circles. The non-interacting $DD^*$ energies (\ref{Eni}) are shown by 
lines: the operators related to black lines are employed, while those related to gray lines are omitted. 
Label [2] in $T_1^{+}(0)$ refers to the multiplicity of non-interacting level $D(1)D^*(1)$. The orange 
stars represent the analytically reconstructed energy levels based on the fitted scattering amplitudes 
and are slightly horizontally shifted for clarity.}
\label{fig:Ecm}
\end{center}
\end{figure}

{\it Interpolators and finite-volume energies:}   In  the non-interacting limit, the $DD^*$ system 
has discrete  energies on a periodic lattice of size $L=N_La$
  \begin{equation}
  \label{Eni}
  E^{\mathrm{ni}}\!=\! E_{D(\vec p_1)}  +E_{D^*(\vec p_2)} ~,\  \vec p_i=\vec n_i\tfrac{2\pi}{L}, \ \vec n_i\in N^3_L 
  \end{equation}  
with $E_{H(\vec p_i)}^{\mathrm{con}}= (m_H^2+\vec p_i^2)^{1/2} $ in the continuum limit. 
The non-interacting energies are shown by lines in Fig. \ref{fig:Ecm}.  
   
The finite-volume energies in the interacting theory are determined from the 
correlation matrices $C_{ij}(t)=\langle O_i(t_{src}\!+\!t)O_j^\dagger (t_{src})\rangle $, 
where $O_i$ refers to operators that annihilate states with the desired quantum numbers.
The  $cc\bar u\bar d$ system is investigated in  inertial frames with total momenta 
$|\vec P|L/(2\pi)=0, 1, \sqrt2, 2$  and finite-volume irreducible representations 
(irreps) in Table \ref{tab:irreps}. These constrain    $DD^*$ scattering in various partial 
waves $l$, of which   $l\!=\!0$ is expected to dominate near threshold. We utilize only meson-meson
interpolators, where each meson is projected to a definite momentum,
\begin{align}
\label{ops}
&O^{DD^*}=\sum_{k,j} A_{kj}~D(\vec p_{1k})D_j^*(\vec p_{2k})~,     \  \vec p_{1k}+\vec p_{2k}=\vec P\\
&=\sum_{k,j} A_{kj}  [(\bar u\Gamma_1 c)_{\vec p_{1k}}(\bar d\Gamma_{2j} c)_{\vec p_{2k}} -  (\bar d \Gamma_1 c)_{\vec p_{1k}}(\bar u\Gamma_{2j} c)_{\vec p_{2k}} ]\nonumber
\end{align}
with two choices  $(\Gamma_1,\Gamma_{2j})\!=\!(\gamma_5,\gamma_j),  (\gamma_5\gamma_t,\gamma_j\gamma_t)$ 
throughout.  
 Operators are shown in Section I of Ref. \cite{Suppl}.
All quark fields are smeared according to the `Distillation' method \cite{HadronSpectrum:2009krc,Piemonte:2019cbi} 
with $60(90)$ Laplacian eigenvectors for $N_L=24(32)$.  
   
The diquark-antidiquark interpolators $[cc][\bar d\bar u]$ are not considered in this work. This is 
justified as it was observed in an earlier lattice calculation that such operators have negligible 
effects on the low-lying energies  \cite{Cheung:2017tnt}. Indications from phenomenological studies 
on the dominance of molecular $DD^*$ Fock components  \cite{Janc:2004qn}  also suggest that $DD^*$ 
interpolators are sufficient to compute the energies faithfully. Furthermore, the application of two 
operators $D_{\gamma_5}D^*_{\gamma_j}$ and $D_{\gamma_5\gamma_t}D^*_{\gamma_j\gamma_t}$ for each momentum 
combination is expected to provide enough variety to extract the energy levels reliably.
      
The  energies $E_n^{\mathrm{lat}}$ are extracted from single-exponential fits to the eigenvalue 
correlators $\lambda^{(n)}(t)\!\propto\! e^{-E_n^{\mathrm{lat}}t}$ of the generalized 
eigenvalue problem $C(t) v^{(n)}(t)= \lambda^{(n)}(t) C(t_0) v^{(n)}(t)$ with $t_0=4$ 
\cite{Michael:1985ne}. In order to mitigate small deviations of single-hadron energies 
$E^{\mathrm{lat}}_{H(\vec p)}$ from $E_{H(\vec p)}^{\mathrm{con}}$ due to discretization effects, 
we take $E_n=E^{\mathrm{lat}}_n+E_{D(\vec p_1)}^{\mathrm{con}}+E_{D^*(\vec p_2)}^{\mathrm{con}}-E_{D(\vec p_1)}^{\mathrm{lat}}-E_{D^*(\vec p_2)}^{\mathrm{lat}}$ 
as the final energies for the scattering analysis, as argued and utilized 
on the same ensembles in Refs. \cite{Piemonte:2019cbi,Prelovsek:2020eiw}.  
      
The resulting finite-volume energies  in the first three irreps are presented in Fig. \ref{fig:Ecm} 
for the heavier charm quark mass. The figure displays the energies $E_{cm}=(E^2-\vec P^2)^{1/2}$ in 
the center-of-momentum frame in units of energy of the $DD^*$ threshold. The large circles and squares 
refer to the energy levels extracted from the lattice simulation. The energy levels have nonzero 
energy shifts with respect to the noninteracting $DD^*$ energies indicating nontrivial interactions. 
These energy shifts render information on the $DD^*$ scattering amplitudes. We find similar observations 
at the lighter charm quark mass \cite{Suppl}.  

\vspace{0.2cm}

\begin{figure}[h!]
\begin{center}
  \includegraphics[width=0.4\textwidth]{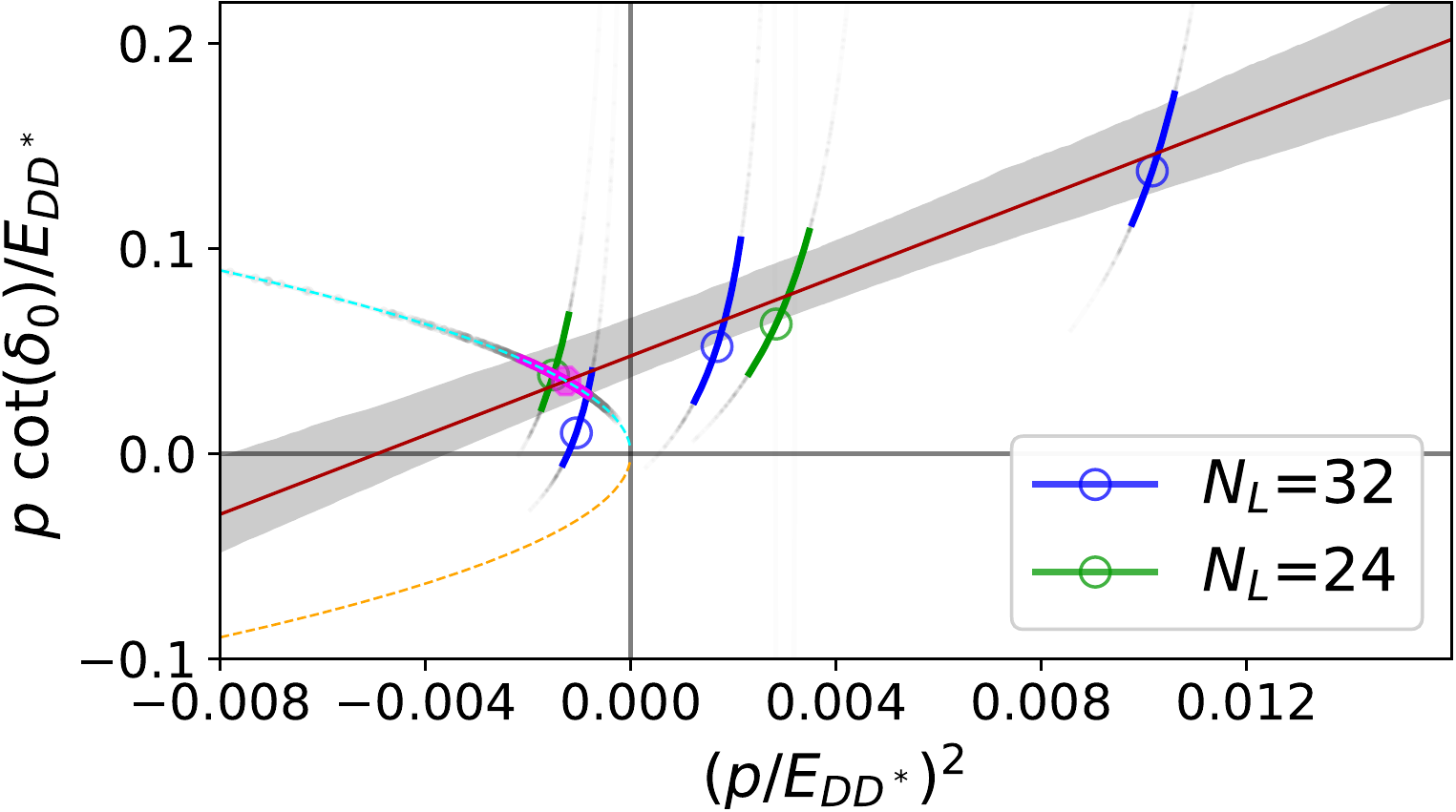}
  \includegraphics[width=0.4\textwidth]{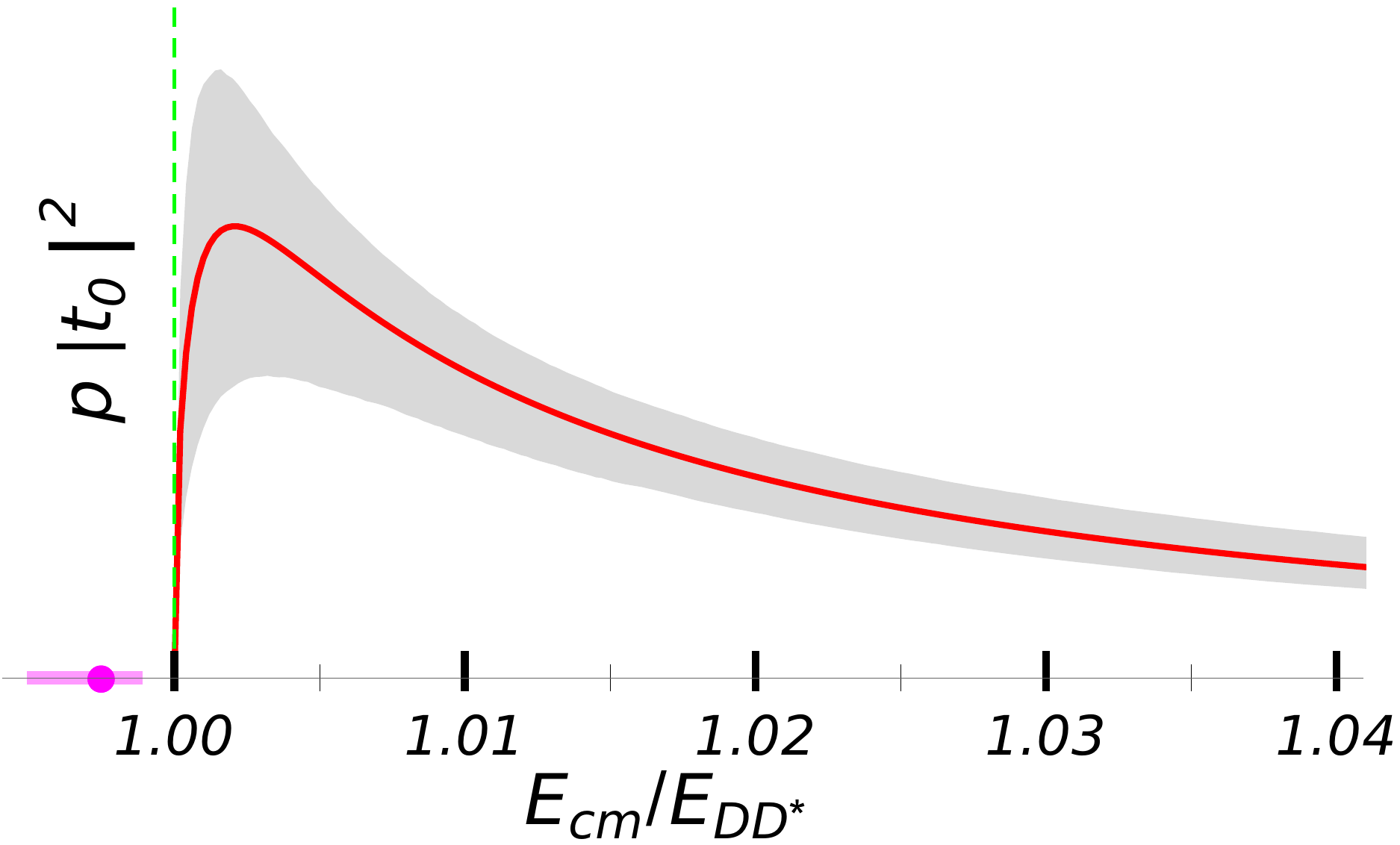}
\caption{Top: $p \cot \delta_{l=0}^{(J=1)}$ for $DD^*$ scattering at the heavier charm quark mass (red line) 
and $ip=+|p|$ (cyan line) versus $p^2$, all normalized to $E_{DD^*}\equiv m_D+m_{D^*}$. The virtual bound 
state occurs at the momenta indicated by the magenta octagon, where two curves intersect. 
Bottom:  The corresponding $DD^*$ scattering rate $ N\propto p |t_0|^2$ above threshold
along with the pole position.
}
\label{fig:pcotdel}
\end{center}
\end{figure}

{\it Scattering analysis:} The scattering amplitude $t$ in $S=e^{2i\delta}=1+i \tfrac{4p}{E_{cm}} t$ 
depends on energy, the partial wave $l$ and $J\!=\!|s-l|,..,|s+l|$, where $s\!=\!1$ for $DD^*$ 
system. We approximate their energy dependence near threshold with two terms of the effective 
range expansion in $p^2$ (\footnote{This relation omits mixing of partial waves for reasons 
discussed later, while the more general relation is provided in Ref. \cite{Suppl}.})
\begin{equation}
\label{t}
t_l^{(J)}= \frac{E_{cm}}{2}\frac{1}{p\cot \delta_{l}^{(J)} - ip}~,\ p^{2l+1} \cot\delta_{l}^{(J)} = \frac{1}{a_l^{(J)}}+\frac{r_l^{(J)}}{2} ~p^2~, 
 % S_l^J=1+i \tfrac{4p}{E_{cm}} ~t_l^J =e^{2i\delta_l^J}~,\ p^{2l+1} \cot\delta_{l}^{J} = \frac{1}{a_l^J}+\frac{1}{2} r_l^J ~p^2~, 
\end{equation}  
where $p=|\vec p|$ is the spatial-momentum of $D$  and $D^*$ in the center-of-momentum frame.
Each finite-volume energy level $E_{cm}$ is related to the $t_l^{(J)}(E_{cm})$ via L\"uscher's 
relation \cite{Luscher:1990ux} and its generalizations, e.g. \cite{Briceno:2014oea}. In order 
to constrain the energy dependence of $t$, the parameters of the effective range expansion are 
optimized such that L\"uscher's relation is simultaneously satisfied for all the energy levels 
considered. For the $l=0$ partial wave, which dominates near threshold, we find
\begin{align}
\label{t-swave}
& p \cot\delta_{l=0}^{(J=1)} = \frac{1}{a_0^{(1)}}+\frac{1}{2} r_0^{(1)} p^2\\
& m_c^{(h)}:\  a_0^{(1)}\!=\!1.04(29)~\mathrm{fm},\  r_0^{(1)}\!=\!0.96(_{-0.20}^{+0.18})~\mathrm{fm}.\nonumber
\end{align}
This fit is shown by the red line in Fig. \ref{fig:pcotdel}. 
  
This result is robust to various fits we have performed, as further detailed in Ref. \cite{Suppl}.  The $J^P\!=\!1^+$ 
is allowed for the $DD^*$ system with spin one in partial waves $l\!=\!0$ and $l\!=\!2$, which 
could lead to a partial wave mixing. We find that $t_{2}^{(1)}$  is consistent with zero, since 
the energy levels with dominant overlaps to $O_{l=2}$  \cite{Suppl} have energies consistent 
with the non-interacting energy (\ref{Eni}). Hence we assume $t_{l\geq 2}^{(J)}=0$ and  
negligible mixing of $l\!=\!2$ with $l\!=\!0$ in $J\!=\!1$ \cite{Suppl}.  The energies 
in blue and green from Fig. \ref{fig:Ecm} are utilized to constrain the energy 
dependence of $t_{0}^{(1)}$ in Eq.~(\ref{t-swave}) and $t_{1}^{(0)}$. We employ a combination 
of procedures outlined in Refs. \cite{Morningstar:2017spu,Woss:2020cmp} in making our fits 
\cite{Suppl}. The fit has $\chi^2/\mathrm{dof} =3.7/5$ and renders the parameters in Eq.~(\ref{t-swave}) 
for $l=0$ scattering and ($a_1^{(0)}=0.076(_{-0.009}^{+0.008})$ fm$^3$, $r_1^{(0)}=6.9(2.1)$ fm$^{-1}$) 
for $l=1$ scattering. The fit results for $t_{1}^{(0)}$ render poles significantly below threshold, 
at energies that are unconstrained by the energy levels, and therefore we do 
not ascribe them any physical significance. The analytically reconstructed energies based on these $t_l^J$
are indicated by orange stars  in Fig. \ref{fig:Ecm} and agree well with the observed energies. 
       
\begin{table*}[tb]
 \begin{center}
  \begin{tabular}{c|ccc| c c| c | c}
   \hline
   & $m_D$~[MeV] & $m_{D^*}$~[MeV] &   $M_{av}$~[MeV] & $a_{l=0}^{(J=1)}$ [fm]  & $r_{l=0}^{(J=1)}$ [fm] & $\delta m_{T_{cc}}$ [MeV] & $T_{cc}$   \\
   \hline
   \hline
   lat.~($m_\pi\! \simeq\! 280~$MeV,$~m_c^{(h)}$) &   1927(1) & 2049(2) &  3103(3)& 1.04(29) & $0.96(_{-0.20}^{+0.18})$  & $-9.9 ^{+3.6}_{-7.2}$ & virtual bound st.\\  
   lat.~($m_\pi\! \simeq\! 280~$MeV,$~m_c^{(l)}$) &   1762(1) & 1898(2) &   2820(3) & 0.86(0.22) & $0.92(_{-0.19}^{+0.17})$ & $-15.0(_{-9.3}^{+4.6})$ & virtual bound st.\\
   exp.~\cite{LHCb:2021auc,pdg2020}  & 1864.85(5) &  2010.26(5) & 3068.6(1) & -7.15(51) & [-11.9(16.9),0]&  $-0.36(4)$ & bound st. \\
   \hline
  \end{tabular}
 \end{center}
\caption{Lattice results for the binding energy $\delta m_{T_{cc}}$ and the effective range parameters in 
Eq. (\ref{t}) at heavier ($m_c^{(h)}$) and lighter ($m_c^{(l)}$) charm quark masses, compared to experiment. 
$~m_c^{(h)}$ is closer to the physical value according to the spin averaged charmonium mass 
$M_{av}\equiv\tfrac{1}{4}(m_{\eta_c}+3m_{J/\psi})$. The real part of experimental $a_0^{(1)}$ is 
provided. The binding energy $\delta m_{T_{cc}}\equiv \mathrm{Re}(E_{cm}^p)-m_{D^0}-m_{D^{*+}}$ is obtained 
from the energy $E_{cm}^p$, where the scattering amplitude has a pole. Lattice results are shown with 
$1\sigma$ statistical errors at given quark masses and lattice spacing; the $T_{cc}$ is found to be a 
virtual bound state with $\delta m_{T_{cc}}<0$ also within $2\sigma$ and $3\sigma$ error ranges.}
\label{tab:results}
\end{table*}

\begin{figure}[h!]
\begin{center}
  \includegraphics[width=0.35\textwidth]{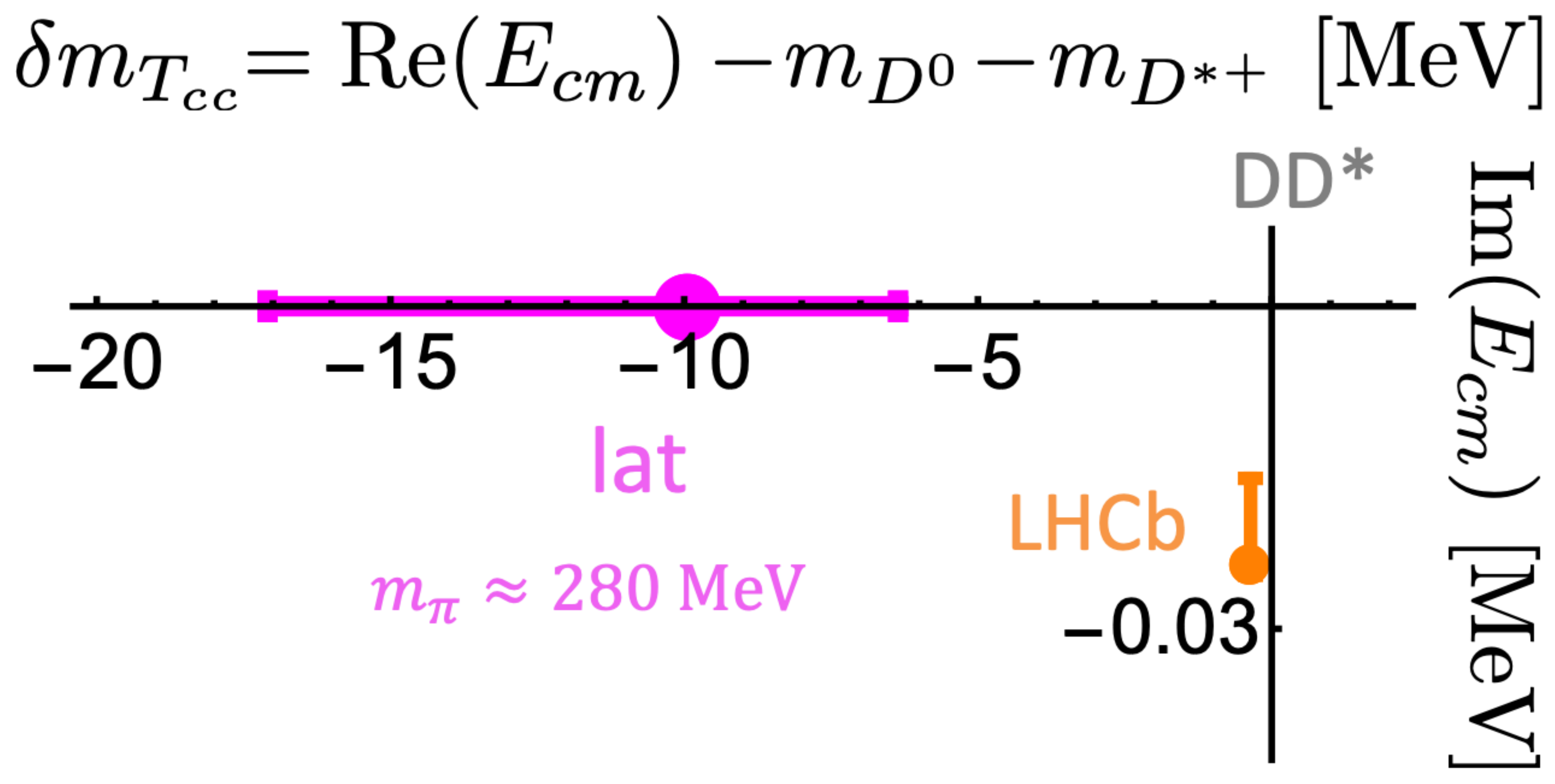}
\caption{The pole in the scattering amplitude related to $T_{cc}$ in the complex energy plane: our lattice 
result at the heavier charm quark mass (magenta) and the LHCb result (orange). 
 }
\label{fig:pole}
\end{center}
\end{figure} 

\begin{figure}[h!]
\begin{center}
\includegraphics[width=0.3\textwidth]{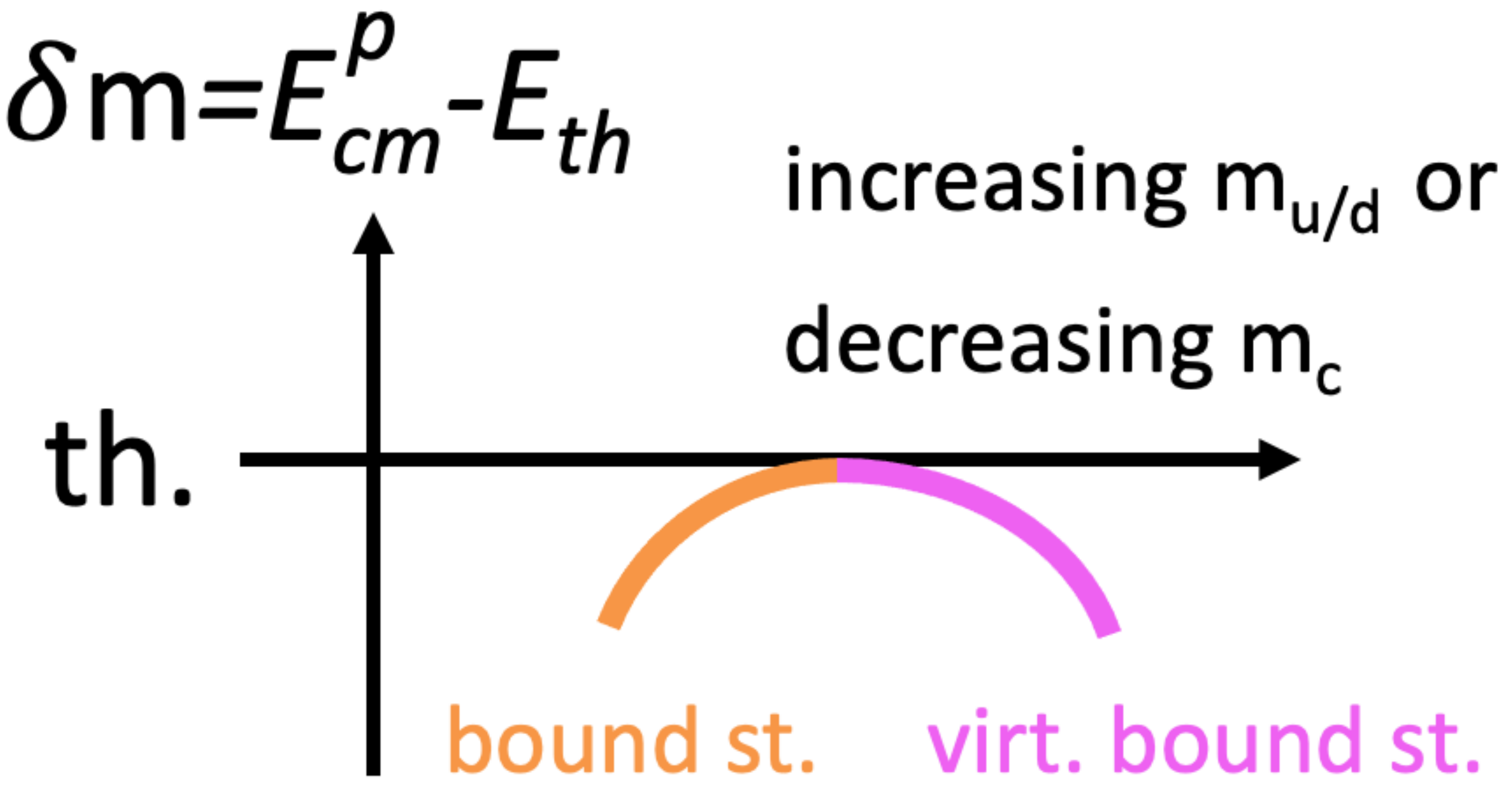}
\caption{Sketch of the binding energy for the (virtual) bound state dominated by the molecular component. 
It is based on a purely attractive potential $V(r)$ and partial wave $l=0$ within quantum mechanics.}
\label{fig:mq-dep}
\end{center}
\end{figure} 

\vspace{0.2cm}

{\it The pole in the $DD^*$ scattering amplitude and $T_{cc}$:} 
Before focusing on $T_{cc}$, let us briefly review the relation between hadrons and poles. 
The existence a hadron state and its mass are inferred from the pole in the scattering amplitude 
$t(E_{cm})$.  The bound state and the  virtual bound state have pole at a real energy below 
threshold and therefore $p^2<0$. A bound state  has pole at $p=i|p_B|$ and is an asymptotic 
state, e.g.   deuteron.  A virtual bound state  has a pole at $p=-i|p_B|$ and is less familiar, 
it appears for example in  $^1S_0$  nucleon-nucleon channel  \cite{Matuschek:2020gqe,Reinert:2017usi,Suppl}.
Finally, the most common poles with $E_{cm}$ away from the real axis correspond to decaying  
resonances, e.g. $\rho$ meson.

We find a virtual bound state pole in $DD^*$ scattering amplitude $t_{l=0}^{(J=1)}$ at energy 
$E_{cm}^p\!=\!(m_D^2-|p_B|^2)^{1/2}\!+\!(m_{D^*}^2-|p_B|^2)^{1/2}$. It corresponds to the binding momentum 
indicated by the magenta octagon in Fig. \ref{fig:pcotdel}. We therefore find an evidence for the doubly 
charmed tetraquark as a virtual bound state with binding energy 
\begin{equation}
\label{mass-tcc}
 m_c^{(h)}:\ \delta m_{T_{cc}}=E_{cm}^p-m_D-m_{D^*}=-9.9_{-7.1}^{+3.6}~\mathrm{MeV}. 
\end{equation}
It is situated slightly below $DD^*$ threshold, close to the mass of the doubly charmed tetraquark $T_{cc}$ 
discovered by LHCb \cite{LHCb:2021vvq,LHCb:2021auc}. The state found on the lattice is strongly stable 
and the pole appears at real energy since $D^*\to D\pi$ is not kinematically allowed for $m_\pi\simeq 280~$MeV. 
The $T_{cc}$ discovered by LHCb decays to $D^0D^0\pi^+$ and the pole is slightly imaginary, as shown in 
Fig. \ref{fig:pole}. The $T_{cc}$ found in experiment would be a bound state in the limit of stable 
$D^{*+}$ since the measured $a_0^{(1)}$ is negative \cite{LHCb:2021auc}.

The quark mass dependence of $T_{cc}$ and the notion of a virtual  bound state can be most easily 
illustrated  for s-wave scattering   in purely attractive potential  $V(r)$ within quantum mechanics. 
Explicit toy-model examples are  given in Refs. \cite{Suppl, Supplv}. The bound state occurs at $p=i|p_B|$, 
its wave function falls as  $e^{ipr}=e^{-|p_B|r}$ outside  the potential and is an asymptotic state.  
As the potential depth is weakened, the bound state  energy  approaches threshold. As the potential is 
weakened even further so that it is not attractive enough to form a bound state, the s-wave bound state 
typically becomes  a virtual bound state.  It occurs at $p=-i|p_B|$ and its wave function  
$e^{ipr}=e^{|p_B|r}$ outside  $V$ is not normalizable, therefore it is not an asymptotic state.  Even so, 
it gives rise to an abrupt enhancement in the scattering cross-section above the threshold when the  pole 
is close below threshold. This enhancement  is shown in Fig. \ref{fig:pcotdel} for $DD^*$ scattering and 
appears due to virtual bound state $T_{cc}$ in our study.

  \vspace{0.1cm}
  
We expect that the virtual bound state pole found in our lattice simulation at unphysical $u/d$ masses 
is related to the $T_{cc}$ discovered by LHCb, as detailed in Section IV of \cite{Suppl}. The would-be 
LHCb bound state is expected to become a virtual bound state with increasing $m_{u/d}$. This is 
sketched in Fig. \ref{fig:mq-dep} for a tetraquark with a significant molecular $DD^*$ component 
attracted by the Yukawa-like potential $V(r)\propto e^{-Mr}/r$, where the mass of the exchanged  
light hadron $M$ increases with increasing $m_{u/d}$.     
     
A near-threshold virtual bound state pole is also observed for the lighter charm quark mass with a 
slightly larger $|\delta m_{T_{cc}}|$, as listed in Table \ref{tab:results}. This observation is 
 consistent with the dependence of pole position on $m_c$ sketched in Fig. \ref{fig:mq-dep}. This arises 
within quantum mechanics via the reduced $DD^*$ mass for purely attractive potential $V(r)$ that is 
assumed to be flavor blind\footnote{Or else, the lattice results for the binding energy at various heavy 
quark masses can be used to examine how good is the heavy flavor symmetry in line with Ref. 
\cite{Baru:2018qkb}.}. %The influence of the virtual bound state pole on the scattering above threshold 
%is diminished as the pole moves further below threshold. 
         
\vspace{0.2cm}

{\it Conclusions:}  We have performed a simulation of $DD^*$ scattering in lattice QCD at 
$m_\pi\simeq 280~$MeV. Unlike other existing lattice investigations in this regard, we extracted 
the near-threshold scattering amplitudes in the flavor channel $cc\bar u\bar d$ with isospin $I=0$. 
Scattering amplitudes for partial waves $l\!=\!0,1$ are determined via the L\"uscher's method, and 
a virtual bound state pole is found for the partial wave $l\!=\!0$. The doubly charm tetraquark with 
$J^P=1^+$ features as a virtual bound state   $9.9_{-7.1}^{+3.6}~$MeV below threshold in 
our simulation with charm quark mass slightly larger than physical. We also observe that the size 
of the binding energy for this virtual bound state increases with decreasing charm quark mass. 

\vspace{0.1cm}

{\it Outlook:} Future lattice studies are desired to reaffirm our findings and inferences.  
The current knowledge could be improved by adding diquark-antidiquark interpolators, exploring 
dependence on quark masses and investigation of discretization effects based on improved actions 
and at smaller lattice spacings.  The simulations at smaller $m_{u/d}$ are required to establish 
whether the pole will approach the $DD^*$ threshold. The simulations at physical $m_{u/d}$ will 
be challenging due to the strong decays $D^*\to D\pi$ and $T_{cc}\to DD\pi$, while the 
formalism is already available in \cite{Blanton:2021mih}.

\vspace{0.5cm}

%{\bf Acknowledgments}
\begin{acknowledgments}
We would particularly like to thank Sara Collins and the members of the RQCD for discussions and 
support related to the computer resources used in this project. We are grateful to J.~J.~Dudek, 
J.~R.~Green, F.-K.~Guo, A.~D.~Hanlon, B. H\"orz, M.~Karliner, L.~Leskovec, M. Mai, N.~Mathur, D.~Mohler, E. Oset, 
S.~Paul, M.~Rosina, M. Sadl, S. Sharpe, and B.-S.~Zou for valuable discussions. We thank our colleagues in CLS 
for the joint effort in the generation of the gauge field ensembles which form a basis for the 
computation. The correlators were computed on the Regensburg Athene2 cluster. We thank the authors 
of Ref. \cite{Morningstar:2017spu} for making the{\it TwoHadronsInBox} package public and C.~B.~Lang 
for contributions to the computing codes we used. S.~P. acknowledges support by Slovenian Research 
Agency ARRS (research core funding No. P1-0035).
\end{acknowledgments}

\bibliographystyle{h-physrev4}
\bibliography{Lgt_Tcc}

% balance columns at end of main text
\onecolumngrid
\clearpage
\twocolumngrid

\begin{center}
  {\large \bf Supplemental material}
\end{center}

% turn on section numbers for the supplement, when it's included in the document
\makeatletter
\c@secnumdepth=4
\makeatother

\newif\ifsepsupp
\sepsuppfalse

\input{supplemental}

\end{document}

%% file: supplemental.tex
\setcounter{equation}{0}
\setcounter{figure}{0}
\setcounter{table}{0}

This supplemental material provides further information on our study of the doubly charm tetraquark channel. 
We present the interpolators relevant for the scattering of a pseudoscalar particle and a vector particle, 
effective energies  and the details on how the scattering amplitudes are extracted. The notion of the virtual 
bound state  is illustrated in quantum mechanics. The dependence of $T_{cc}$ pole on the quark masses is 
investigated based on simple quantum mechanical arguments.

\maketitle

\section{Interpolators}
 
This section presents the explicit expressions for two-meson interpolators that transform 
according to the irreducible representations $\Lambda$ in Table 1 of the main article. They 
are relevant for the scattering of a pseudoscalar meson $P$ and a vector meson $V$, so 
they are valuable for $DD^*$ scattering simulated in this work and also for many interesting 
channels like $BB^*$, $\pi J/\psi$, $K D^*$ etc. Each meson is projected to a definite momentum, 
which is given in units of $2\pi/L$ in parenthesis.  The linear combinations of momenta and 
vector-meson polarizations are chosen such that the operators transform according to the 
finite-volume irreps. They are  obtained with the partial-wave method for total momentum 
$\vec P\!=\! \vec 0$ \cite{Prelovsek:2016iyo}. For $\vec P\not = \vec 0$ we consider only 
the one-dimensional irreducible representations and the operators are obtained with 
the projection method as $O=\sum_{R\in LG} \chi^\Lambda (R)~ R~P(\vec p_1)V_k(\vec p_2)~ R^{-1}$, 
where $\chi^\Lambda (R)$ is the character. The operators indicated by $O$ were analyzed in 
the present simulation, while operators indicated by $O^\prime$ were not implemented and 
may be valuable for future studies:
{\small
 \begin{align*}
 \bm{T_1^+},& \  \bm{\vec P=\{0,0,0\}}  ,\  \mathrm{row\ z}  \nonumber\\ 
 O^{l=0}= &P(\{0,0,0\}) V_z(\{0,0,0\})   \\ 
  O^{l=0}= &P(\{1,0,0\})  V_z(\{-1,0,0\})+P(\{-1,0,0\}) V_z(\{1,0,0\})\nonumber\\
  +&P(\{0,1,0\}) V_z(\{0,-1,0\})+P(\{0,-1,0\}) V_z(\{0,1,0\}) \nonumber\\
  +&P(\{0,0,1\})V_z(\{0,0,-1\})+P(\{0,0,-1\}) V_z(\{0,0,1\}) \bigr]    \nonumber\\
   O^{l=2} =  & P(\{1,0,0\}) V_z(\{-1,0,0\})+P(\{-1,0,0\}) V_z(\{1,0,0\})\nonumber\\
  +&P(\{0,1,0\}) V_z(\{0,-1,0\})+P(\{0,-1,0\}) V_z(\{0,1,0\}) \nonumber\\
  -2[&P(\{0,0,1\})V_z(\{0,0,-1\})+P(\{0,0,-1\}) V_z(\{0,0,1\})\bigr] \nonumber\\ 
O^{l=0}=&V_{1x}[{0, 0, 0}] V_{2_y}[{0, 0, 0}] - V_{1y}[{0, 0, 0}] V_{2x}[{0, 0, 0}]   \nonumber\\
&~\\
 \bm{A_1^-},& \  \bm{\vec P=\{0,0,0\}}   \nonumber\\ 
 O=&P(\{1,0,0\}) V_x(\{-1,0,0\})-P(\{-1,0,0\}) V_x(\{1,0,0\})\\
 +&P(\{0,1,0\}) V_y(\{0,-1,0\})-P(\{0,-1,0\}) V_y(\{0,1,0\})\nonumber\\
 +&P(\{0,0,1\})  V_z(\{0,0,-1\})-P(\{0,0,-1\}) V_z(\{0,0,1\}) 
 \end{align*}
 \begin{align*}
 \bm{  A_2},& \ \bm{\vec P=\{0,0,1\}}   \nonumber\\ 
 O=&P(\{0,0,0\}) V_z(\{0,0,1\}) \nonumber\\  
       O=&P(\{0,0,1\}) V_z(\{0,0,0\})  \nonumber\\
    O^\prime=&P(\{1,0,1\}) V_x(\{-1,0,0\})-P(\{-1,0,1\}) V_x(\{1,0,0\})\nonumber\\
                   +&P(\{0,1,1\}) V_y(\{0,-1,0\})-P(\{0,-1,1\}) V_y(\{0,1,0\})  \nonumber\\
    O^\prime=&P(\{1,0,1\}) V_z(\{-1,0,0\})+P(\{-1,0,1\}) V_z(\{1,0,0\})\nonumber\\
                   +&P(\{0,1,1\})  V_z(\{0,-1,0\})+P(\{0,-1,1\})  V_z(\{0,1,0\}) \nonumber\\
    O^\prime=&P(\{1,0,0\}) V_x(\{-1,0,1\})-P(\{-1,0,0\})  V_x(\{1,0,1\})\nonumber\\
                  +&P(\{0,1,0\})  V_y(\{0,-1,1\})-P(\{0,-1,0\}) V_y(\{0,1,1\})    \nonumber\\
    O^\prime=&P(\{1,0,0\})  V_z(\{-1,0,1\})+P(\{-1,0,0\})  V_z(\{1,0,1\})\nonumber\\
                  +&P(\{0,1,0\}) V_z(\{0,-1,1\})+P(\{0,-1,0\}) V_z(\{0,1,1\}) \nonumber    
 \end{align*}
 \begin{align*}
 \bm{  A_2},& \ \bm{\vec P=\{1,1,0\}}   \nonumber\\                
  O=& P(\{0,0,0\}) (V_x(\{1,1,0\})+ V_y(\{1,1,0\}))  \\
   O=&P(\{1,0,0\}) V_x(\{0,1,0\})+P(\{0,1,0\})  V_y(\{1,0,0\})  )\nonumber\\
   O=&P(\{0,1,0\})  V_x(\{1,0,0\})+P(\{1,0,0\})  V_y(\{0,1,0\}) \nonumber\\
   O=&P(\{1,1,0\}) ( V_x(\{0,0,0\})+ V_y(\{0,0,0\})) \nonumber\\ 
     &~\\
 \bm{  A_2},& \ \bm{\vec P=\{0,0,2\}}   \nonumber\\ 
               O=&P(\{0,0,1\})  V_z(\{0,0,1\})  \\       
    O^\prime=&P(\{0,0,0\})  V_z(\{0,0,2\})     \nonumber\\ 	
    O^\prime=&P(\{0,0,2\})  V_z(\{0,0,0\})
  \end{align*}
}
The number of pseudoscalar-vector eigen-states is equal to the number of interpolators 
in the non-interacting limit. This renders degenerate eigenstates in non-interacting 
limit that are indicated by [2] in Fig. 1 of the main article. This is responsible 
for nearly degenerate states in interacting theory - we observe all the expected nearly 
degenerate states in our finite volume energy levels. The current study also employs 
the $D^*D^*$ interpolator for the $T_1^+$ irrep and considers the levels below $D^*D^*$
threshold in the scattering analysis.    
  
  \newpage
   
  \section{Finite-volume energies and effective energies  }
  
\begin{figure}[tbh!]
\begin{center}
\includegraphics[width=0.49\textwidth]{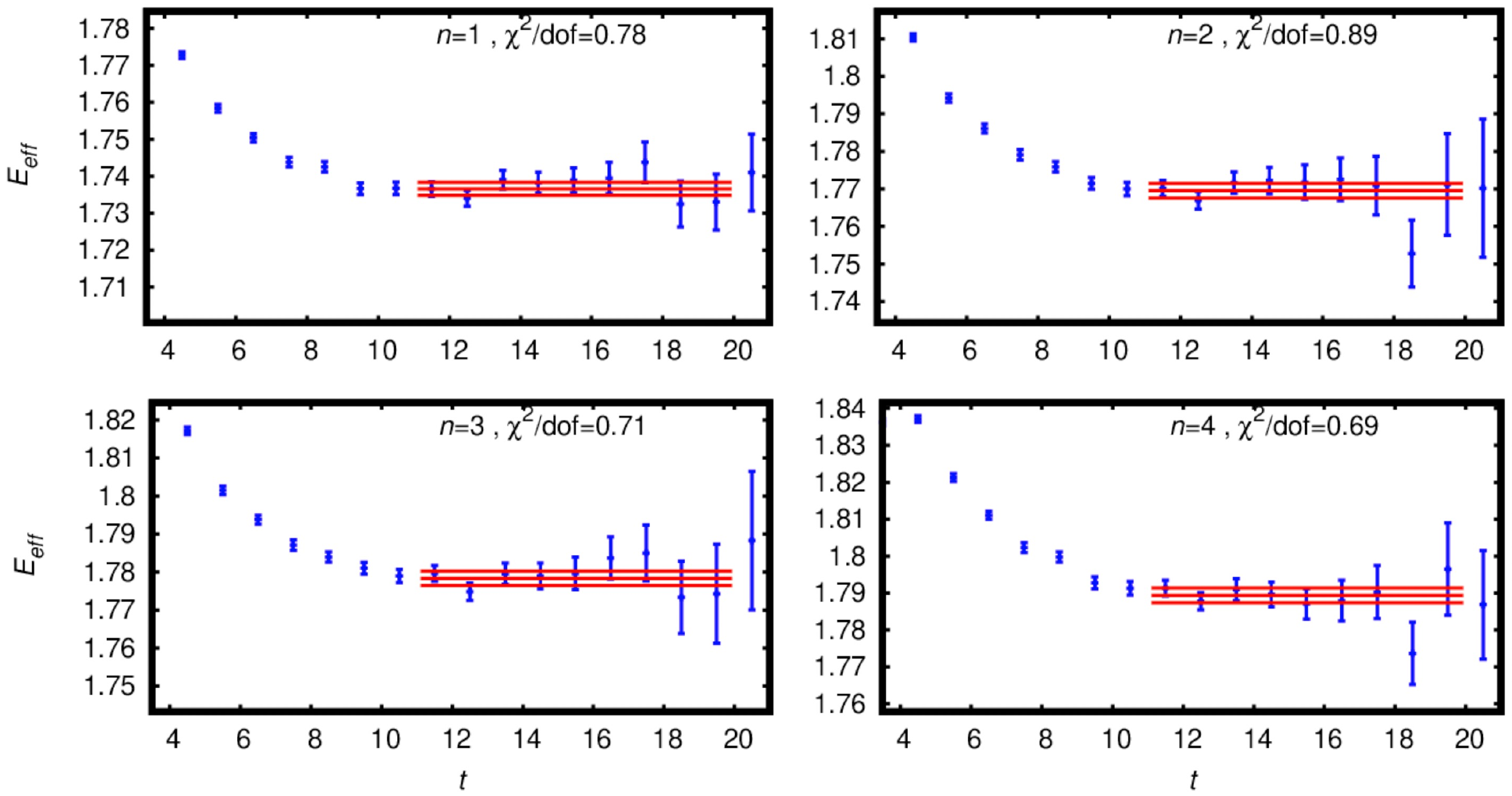}
\caption{Effective energies $E_{\text{eff}}$ for irreducible representation $T_1^+$ and $N_L=32$. }\label{fig:Eeff}
\end{center}
\end{figure}

\begin{figure*}[tbh!]
\begin{center}
\includegraphics[height=9cm,width=15cm]{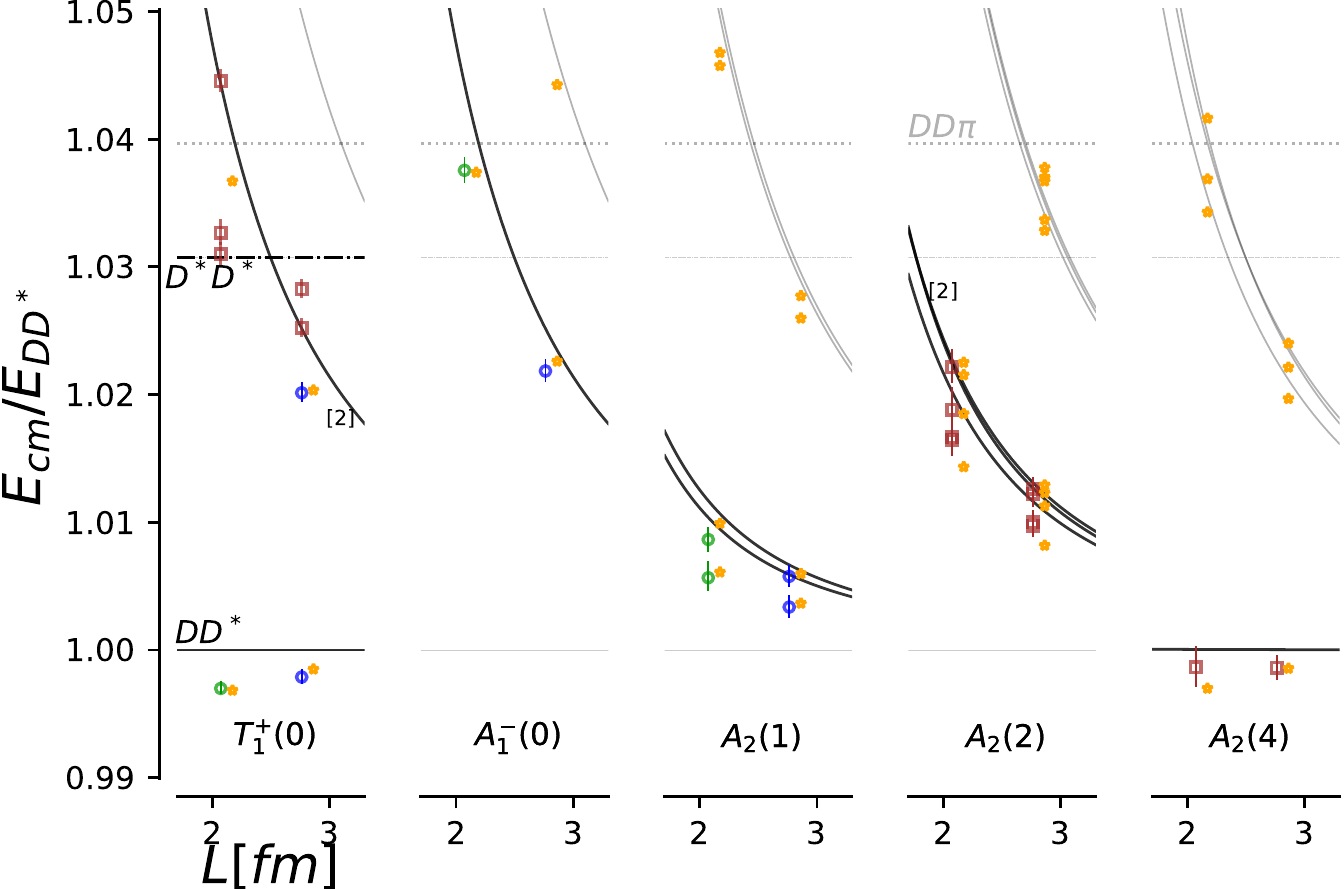}
\caption{Finite-volume energy levels ($E_{cm}$) in the center-of-momentum frame in units of $E_{DD^*}\equiv m_D+m_{D^*}$
for the heavier charm quark mass $m_c^{(h)}$. The results from lattice simulation are shown by empty circles
and squares: blue and green circles are employed in the extraction of $DD^*$ scattering amplitude with $l=0,1$. 
The non-interacting $DD^*$ energies (Eq. (1) in the main text) are shown by lines: the operators related 
to black lines are considered, while the those related to the gray lines are omitted from the calculation. 
[2] in $T_1^{+}(0)$ and $A_2(2)$ refers to the multiplicity of non-interacting level $D(1)D^*(1)$, in 
these irreps. The orange stars represent the analytically predicted energies based on the fitted scattering 
amplitudes and are slightly horizontally shifted for clarity. }
\label{fig:Ecm1}
\end{center}
\end{figure*}

\begin{figure*}[tbh!]
\begin{center}
\includegraphics[height=9cm,width=15cm]{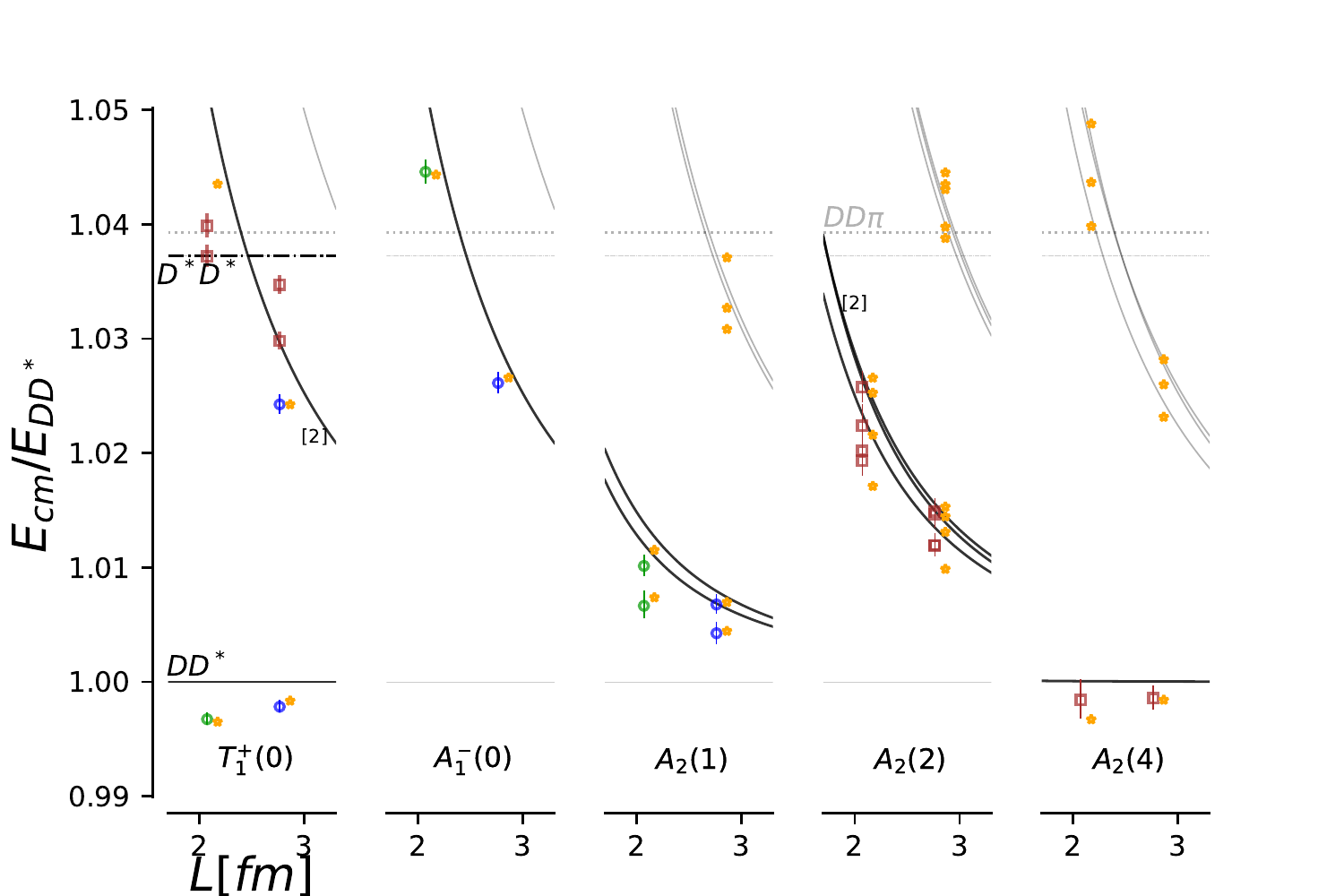}
\caption{Same as in Fig. \ref{fig:Ecm1}, but for the lighter charm quark mass $m_c^{(l)}$.}
\label{fig:Ecm2}
\end{center}
\end{figure*}

Examples of effective energies are shown in Fig. \ref{fig:Eeff} for  the irreducible representation 
$T_1^+$ and $N_L=32$. Energy estimates from single exponential fits  in the plateaued regions are 
indicated by red horizontal lines.    The resulting finite-volume energies for all five irreducible 
representations and both volumes $N_L=24,32$  are shown in Figs.  \ref{fig:Ecm1} and \ref{fig:Ecm2} 
for two charm quark masses, respectively. 

These results are obtained from the  correlation matrices that are averaged 
over all spin and momentum polarizations and over several source timeslices $t_{src}$.

 \section{Details of scattering analysis} 
 
In this section, we discuss the details of our procedure for extracting the scattering 
amplitude $t$ in Eq. (4) from the finite-volume energies and present a summary of various 
fits we have performed. The best fit values of parameters in the effective range 
expansion  describing the energy dependence of $t$ are determined by minimizing 
a $\chi^2$ function defined as 
\begin{align}
\chi^2(\{a\}) =& \sum_{L} \sum_{\vec{P}\Lambda n}\sum_{\vec{P}'\Lambda' n'} dE_{cm}(L, \vec{P}\Lambda n; \{a\})\\ 
               & \mathcal{C}^{-1}(L; \vec{P}\Lambda n; \vec{P}'\Lambda' n') dE_{cm}(L, \vec{P}'\Lambda' n'; \{a\})~.\nonumber 
\label{chisq}
\end{align}
Here 
\begin{equation}
dE_{cm}(L, \vec{P}\Lambda n; \{a\}) = E_{cm}(L, \vec{P}\Lambda n) - E_{cm}^{an.}(L, \vec{P}\Lambda n; \{a\}) \nonumber 
\end{equation}
is the difference between  an observed lattice energy level $E_{cm}(L, \vec{P}\Lambda n)$ and 
the analytically calculated energy level $E_{cm}^{an.}(L, \vec{P}\Lambda n; \{a\})$ that satisfies the generalized 
L\"uscher's equation \cite{Briceno:2014oea, Morningstar:2017spu} 
\begin{equation}
\label{QCn}
\det[ (\tilde K^{(J)}_{ls;l's'}(E_{cm},\{a\}))^{-1}\delta_{JJ'} -\delta_{ss'}B^{\vec P,\Lambda}_{lJ;l^\prime J^\prime}(E_{cm})]=0
\end{equation}
for a given set of parameter values $\{a\}$. 
The $B$ is a known kinematical matrix, computed using the {\it TwoHadronsInBox} package \cite{Morningstar:2017spu} 
and $\tilde K^{-1}(E_{cm},\{a\})$ is related to $t$ as
\begin{equation}
\label{Rkn0}
(t^{(J)}_{ls;l's'})^{-1} = \frac{2(\tilde K^{(J)}_{ls,l's'}(E_{cm},\{a\})^{-1}}{E_{cm}p^{l}p^{l'}} - i \frac{2p\delta_{ll'}\delta_{ss'}}{E_{cm}}.
\end{equation}
Here ($l$, $s$, $J$) refer to the partial-wave, the total spin and the total angular momentum of the incoming 
particles involved in the scattering, whereas the primed variables refer to that of the outgoing particles. 
The data covariance $\mathcal{C}(L; \vec{P}\Lambda n; \vec{P}'\Lambda' n')$ is determined using the procedure 
outlined in Appendix A of Ref. \cite{Prelovsek:2020eiw}. The solutions of the L\"uscher's equation (\ref{QCn}) 
are extracted from the zeros in eigenvalues (as a function of $E_{cm}$ for each lattice QCD ensemble and 
finite-volume irrep) of the matrix
\begin{equation}
\label{Omega}
\tilde A(E_{cm}) = \frac{A}{\det((\mu^2+AA^{\dagger})^{1/2})},
\end{equation}
by performing an eigenvalue decomposition along the lines as discussed in Ref. \cite{Woss:2020cmp}.
Here $A$ is the argument of determinant in Eqn. \ref{QCn} and $\mu=2.0$ is chosen throughout the calculation. 
We find our results are independent of the value of $\mu$ across a wide interval [0.1, 100.].

Under the assumption that any contribution from $l\ge2$ partial waves are negligible\footnote{All 
the energy levels with dominant overlap to operators with $l=2$ partial waves are consistent with 
the respective non-interacting energies. Any contribution of $l\ge2$ partial waves on the finite 
volume levels close to the $DD^*$ threshold, which is the energy region of interest, are 
suppressed by the phase space factor $p^{2l}$. The extraction of any other effects from $l\ge2$ 
partial waves need the utilization of larger volume ensembles, which is beyond the scope of this 
work.}, $\tilde K^{-1}(E_{cm},\{a\})$ for elastic $DD^*$ scattering reduces to a $3\times3$ 
diagonal matrix, where elements are related to $t$ and $\delta$  as follows
\begin{equation}
\label{Rkn}
(t_l^{(J)})^{-1} = \frac{2(\tilde K_l^{(J)})^{-1}}{E_{cm}p^{2l}} - i \frac{2p}{E_{cm}}, \quad  (\tilde K_l^{(J)})^{-1} =p^{2l+1} \cot \delta_l^{(J)}
\end{equation}
We parametrize it  with the effective range expansion 
\begin{equation}
\label{eq:Kinv}
\tilde K^{-1}=
   \left[
 \begin{array}{ccc}
  \frac1{a_0^{(1)}}+\frac{r_0^{(1)}p^2}2 & 0 & 0 \\
  0 & \frac1{a_1^{(0)}}+\frac{r_1^{(0)}p^2}2 & 0 \\
  0 & 0 & \frac1{a_1^{(2)}}
\end{array}
\right].
\end{equation}

\begin{figure}[h!]
\begin{center}
\includegraphics[width=0.4\textwidth]{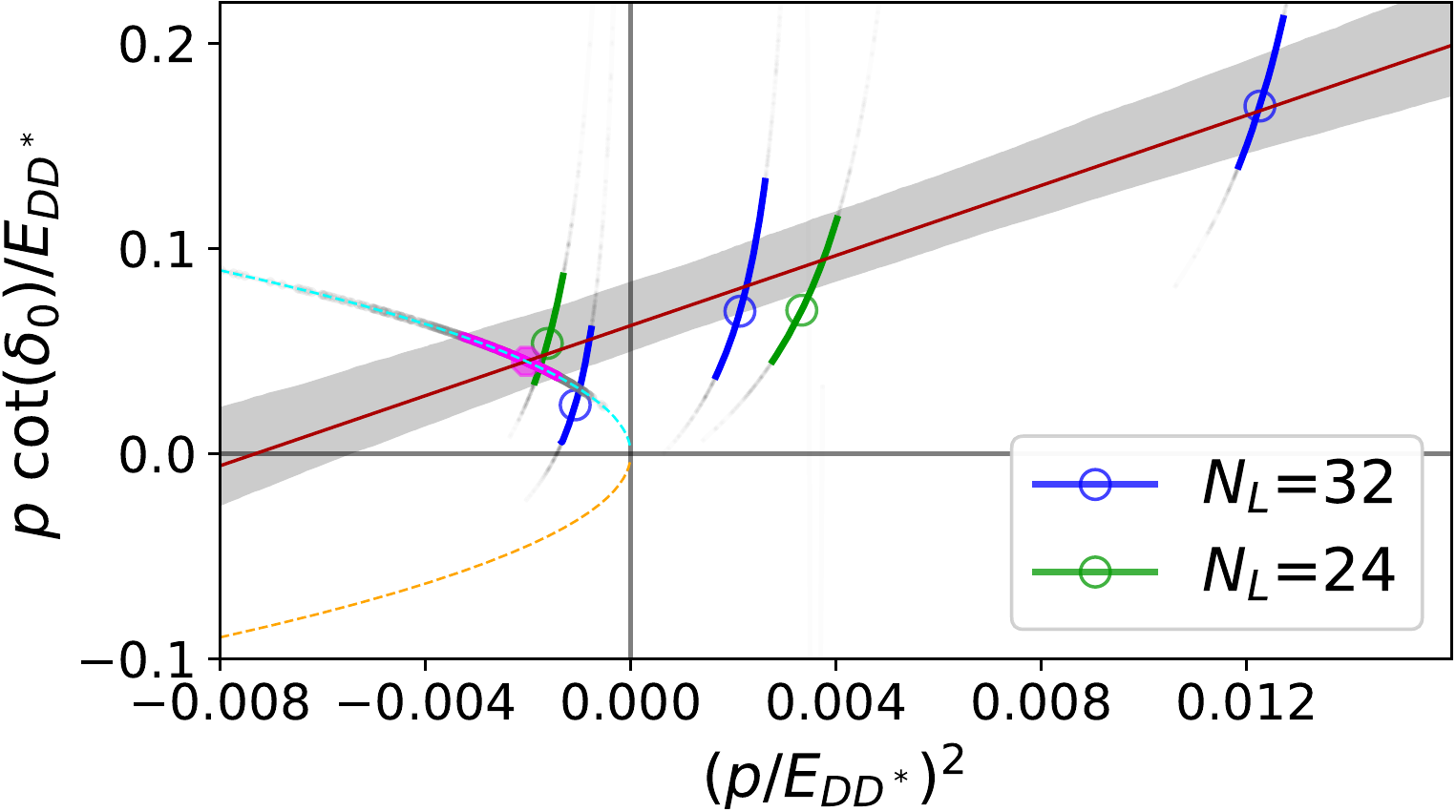}
\caption{$p \cot \delta_{l=0}^{(J=1)}$ versus $p^2$ in units of $E_{DD^*}\equiv m_D+m_{D^*}$ for 
$DD^*$ scattering at the light charm quark mass. The (cyan) orange dashed curve refers to the 
(virtual) bound state constraint. Green and blue circles indicate the simulated data, whereas
the solid red curve along with the gray band are the fit results. The virtual bound state occurs at 
the momenta indicated by the magenta octagon, where red and cyan curves intersect. }
\label{fig:pcotdel2}
\end{center}
\end{figure}

We perform separate and combined fits to the $J^P(l)=1^+(0), 0^-(1)$ channels (first two rows in 
the above $\tilde K^{-1}$ matrix) using $T_1^+$ and $A_1^-$ irreps in the rest frame and $A_2$ irrep
in the moving frame with $\vec P= (0,0,1)\frac{2\pi}L$. Possible effects from the left-hand cuts 
are omitted. The fitting details, quality, results, and the parameter covariance for all fits in 
the two $m_c$ we studied are listed in Table \ref{tab:fitresults}. It is evident that the best fit 
parameters for $l=0$ $DD^*$ scattering amplitude is stable with respect to separate and combined 
fits in the low energy region. $p \cot(\delta_0)$ as a function of $p^2$ is presented in Fig. 2 
of the main text and Fig. \ref{fig:pcotdel2} below for the $m_c^{(h)}$ and $m_c^{(l)}$, respectively. 

\begin{table*}[tbh!]
 \begin{center}
  \begin{tabular}{c|c|c|c|l|c|c|rrrr|c|c}
   \hline
   $m_c$ & ID & $l$ &  $\Lambda(|\vec P|^2)$ & $E^{lat}$ info & $\chi^2$/dof & $\{a\}$ & \multicolumn{4}{c|}{$\{a\}$ Covariance} & $\delta m_{T_{cc}}~$[MeV] & $\bar X_A$ \\ 
         &     & &                       & $N_L=(24, 32)$ &              &                &      &         &       &     &  &  \\ \hline
   $m_c^{(h)}$ & 1 & 0   & $T_1^+(0)$ & (1000, 1100) & 1.3/3 & $a_0^{(1)}=1.13(_{-0.34}^{+0.37})$ fm & 1.00 & 0.06 &       &        & \multirow{2}{*}{$-8.9(_{-7.0}^{+3.2})$} & \multirow{2}{*}{0.61(7)} \\ 
               &   &     & $A_2(1)$   & (10, 10)     &       & $r_0^{(1)}=0.94(_{-0.20}^{+0.19})$ fm &      & 1.00 &       &        & & \\ 
   \cline{2-13}                                                                                                                   
               & 2 & 0,1 & $T_1^+(0)$ & (1000, 1100) & 3.7/5 & $a_0^{(1)}=1.04(0.29)$ fm             & 1.00 & 0.13 & -0.28 & -0.18  & \multirow{2}{*}{$-9.9(_{-7.2}^{+3.6})$} &  \multirow{2}{*}{0.59(6)} \\ 
               &   &     & $A_2(1)$   & (11, 11)     &       & $r_0^{(1)}=0.96(_{-0.20}^{+0.18})$ fm &      & 1.00 &  0.02 &  0.02  & & \\ 
               &   &     & $A_1^-(0)$ & (1, 1)       &       & $a_1^{(0)}=0.076(_{-0.009}^{+0.008})$ fm$^3$ &      &      &  1.00 &  0.65  & & \\
               &   &     &            &              &       & $r_1^{(0)}=6.9(2.1)$ fm$^{-1}$               &      &      &       &  1.00  & & \\ \hline
   $m_c^{(l)}$ & 3 & 0   & $T_1^+(0)$ & (1000, 1100) & 1.4/3 & $a_0^{(1)}=0.94(0.25)$ fm             & 1.00 & 0.06 &       &        & \multirow{2}{*}{$-12.9(_{-8.2}^{+4.1})$} &  \multirow{2}{*}{0.57(6)} \\ 
               &   &     & $A_2(1)$   & (10, 10)     &       & $r_0^{(1)}=0.96(_{-0.20}^{+0.18})$ fm &      & 1.00 &       &        & & \\ 
   \cline{2-13}                                                                                                                   
               & 4 & 0,1 & $T_1^+(0)$ & (1000, 1100) & 3.6/5 & $a_0^{(1)}=0.86(0.22)$ fm             & 1.00 & 0.12 &  0.09 & -0.11  & \multirow{2}{*}{$-15.0(_{-9.3}^{+4.6})$} &  \multirow{2}{*}{0.56(5)} \\ 
               &   &     & $A_2(1)$   & (11, 11)     &       & $r_0^{(1)}=0.92(_{-0.19}^{+0.17})$ fm   &      & 1.00 & -0.03 &  0.04  & & \\ 
               &   &     & $A_1^-(0)$ & (1, 1)       &       & $a_1^{(0)}=0.117(_{-0.014}^{+0.013})$ fm$^3$  &      &      &  1.00 & -0.95  & & \\
               &   &     &            &              &       & $r_1^{(0)}=8.6(_{-1.1}^{+1.5})$ fm$^{-1}$ &      &      &       &  1.00  & & \\ \hline
  \end{tabular}
 \end{center}
\caption{Details and results of the scattering analysis: the partial waves considered, the 
finite-volume energies included, the quality, the best fit values and the covariances for the 
resulting parameters of various fits. In the last two columns, we also present the binding energy 
and a compositeness measure, as defined in Ref. \cite{Matuschek:2020gqe}, of the $l=0$ virtual bound state.}
\label{tab:fitresults}
\end{table*}

In order to affirm our findings, we analytically predict the finite-volume energies in all five 
irreps. For this, we assume the full form of the $\tilde K^{-1}$ matrix in Eq. (\ref{eq:Kinv}) 
taking fit results with ID (2) and (4) in Table \ref{tab:fitresults} and a small constant value 
for $a_1^{(2)}$ to imitate a non-interacting scenario in this channel. This is then compared with 
the simulated finite volume energies to assess the quality of prediction. The $\chi^2/dof$ 
evaluated using the predicted energies and the simulated results are $23/14$ and $30/14$ 
for the $m_c^{(h)}$ and $m_c^{(l)}$, and are presented in Figs. \ref{fig:Ecm1} and 
\ref{fig:Ecm2}, respectively. We find that the pattern of predicted and simulated finite 
volume energies are consistent, as shown by the smaller and larger symbols in these figures. 
We also observe that expanding the $\tilde K^{-1}$ matrix to include nearly non-interacting 
$J^P(l)=1^+(2), 2^+(2)$ channels, also involving a possible mild partial wave mixing between 
$J^P(l)=1^+(0)$ and $J^P(l)=1^+(2)$ does not alter the finite-volume energies up to 
$E_{cm}/E_{DD^*}=1.025$ and also leads to the same $\chi^2/\mathrm{dof}$ value. In short, we 
find that our estimate for $DD^*$ scattering amplitude  with $l=0$ is robust to contaminations 
from other possible nearby channels.

\section{Bound states, virtual bound states and their
dependence  on  quark masses }

Here we provide a simple quantum mechanical illustration of a virtual bound state and a more 
familiar bound state. Then we argue that one state can convert to the other  as the quark masses 
change, and we explore their binding energies. The aim is to argue that the $T_{cc}$ pole determined 
from lattice simulation and from the experiment roughly varies with changing $m_{u/d}$ or $m_c$ as 
sketched in Fig. 4 of the main article and hence are related.  
  
 \subsection{Bound states and virtual bound states in square well potential}
 
 \begin{figure*}[tbh!]
\begin{center}
\includegraphics[width=0.99\textwidth]{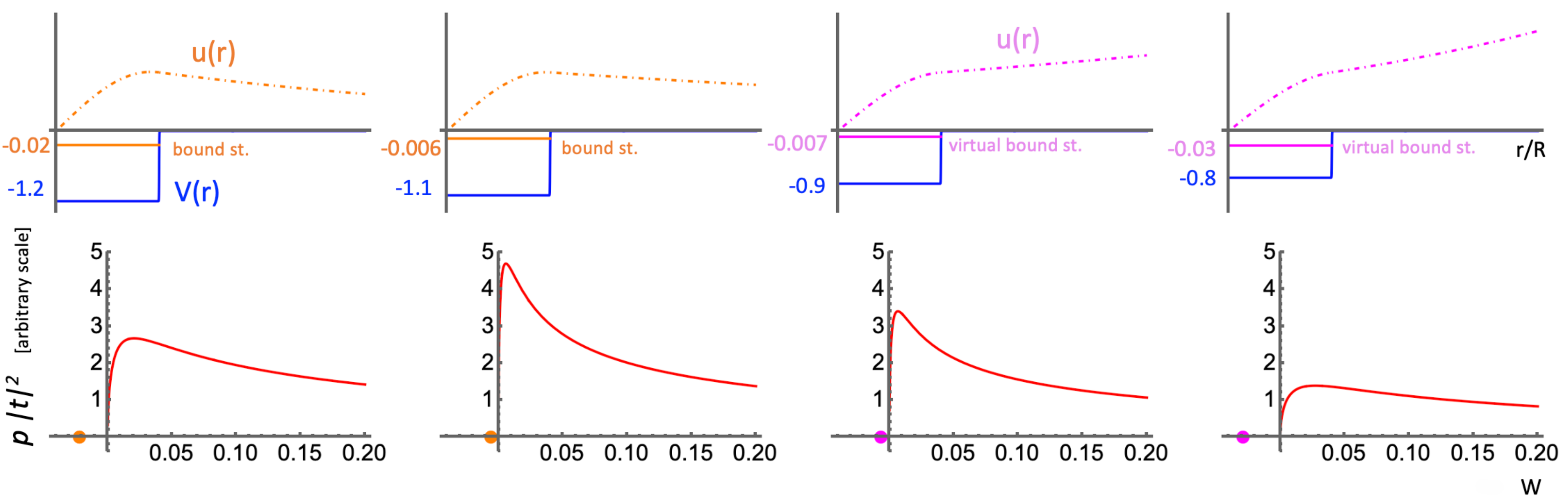}
\caption{  Dependence of various observables related to the scattering in the square-well potential 
as the attraction potential $V_0$ (blue) is decreased, while   $R=1$ and $m_r=\pi^2/8$ are fixed.  
The bound state (orange) becomes a virtual bound state (magenta) at $V_0=1$. Top:  Energy  $W\!<\!0$ 
where the scattering amplitude has a bound state pole (solid orange) or a virtual bound state pole 
(solid  magenta), as well as  the corresponding wave functions at these energies (dot-dashed). 
Bottom: The   scattering rates $N\propto p|t|^2$ above threshold (red) are enhanced due to the poles 
closely below threshold (orange/magenta circles).} 
\label{fig:square-well}
\end{center}
\end{figure*}
 
The notion of a virtual bound state can be most easily illustrated for the case of the square well 
attractive potential $V(r)=-V_0 \theta(R-r)$ between two particles. The scattering phase shift for 
partial wave $l=0$ is obtained by matching the values and derivatives of the  wave functions inside 
$u(r)=A \sin(qr)$ and outside $u(r)=B\sin(pr+\delta)$ the potential at $r=R$, rendering 
$\delta_{0}=\arctan[\tan(qR)\tfrac{p}{q}]-pR$.  Here $p=(2m_rW)^{1/2}$, $q=(2m_r[W+V_0])^{1/2}$, 
$m_r$ is the reduced mass, and $W$ is the non-relativistic energy. Defining the scattering amplitude 
$t$ in the non-relativistic theory as $S=e^{2i\delta_0}=1+2ipt$, one can extract $t$ and determine 
$W$ and $p$ where $t$ has a pole.    
   
The pole positions and the corresponding wave functions are sketched for several values of the attraction 
$V_0$ in Fig. \ref{fig:square-well} (top row). The poles appear at $W<0$ and therefore at imaginary 
momenta $p=\pm i|p_B|$.  The bound state is defined as a state with pole at  $p=+ i|p_B|$ when the wave 
function falls exponentially $e^{ipr}=e^{-|p_B|r}$, and it is present for large attraction $V_0$. As 
$V_0$ is decreased, the binding energy of the bound state decreases.  As $V_0$ is reduced further, 
the bound state turns to virtual bound state, which is defined as a state with a pole at $p=-i|p_B|$. 
Its wave function exponentially increases $e^{ipr}=e^{|p_B|r}$ therefore it is not a normalizable quantum 
mechanical state. Such a state is understood to be a result of weakly attractive potential  between the 
scattering particles, where the interaction is not attractive enough to form a bound state. However, 
the presence of its pole closely below threshold still significantly enhances the cross-section and rate 
$N\propto p|t|^2$ at energies above threshold, as shown in Fig. \ref{fig:square-well} (bottom row). 
Analogous behavior is observed for other shapes of attractive potential, as demonstrated for 
the attractive Gaussian potential in video \cite{Supplv}. 
   
An example of a virtual bound state features in $pp,~pn$ and $nn$ scattering in the channel $^1S_0$ 
with $I\!=\!1$,  where the measured effective parameters are $a_0^{pn}\simeq 23.7~$fm and 
$r_0^{pn}\simeq 2.7~$fm \cite{Matuschek:2020gqe,Reinert:2017usi}. This renders a virtual bound state 
about $66~$keV below threshold and a significant peak in cross section above it.  
      
The dependence of observables for decreasing attraction $V_0$ is shown in Fig. \ref{fig:square-well} 
for fixed $m_r$ and $R$. Similar behavior is observed when $m_r$ is decreased or $R$ is decreased 
(while keeping the other two parameters fixed).

\subsection{Dependence of $T_{cc}$ binding energy on $m_{u/d}$ or $m_c$}
 
In this section we provide some simple quantum mechanic arguments that the pole position of $DD^*$ scattering 
in partial wave $l\!=\!0$ roughly varies with changing $m_{u/d}$ or $m_c$ as sketched in Fig. 4 of the main 
article. We consider a pole related to the state dominated by a {\it molecular Fock component}. The pole 
positions found by LHCb and by our lattice study at two charm quark masses are in line with these arguments. 
However, further lattice studies at various quark masses are needed to establish these arguments empirically. 
      
\begin{figure}[tbh!]
\begin{center}
\includegraphics[width=0.25\textwidth]{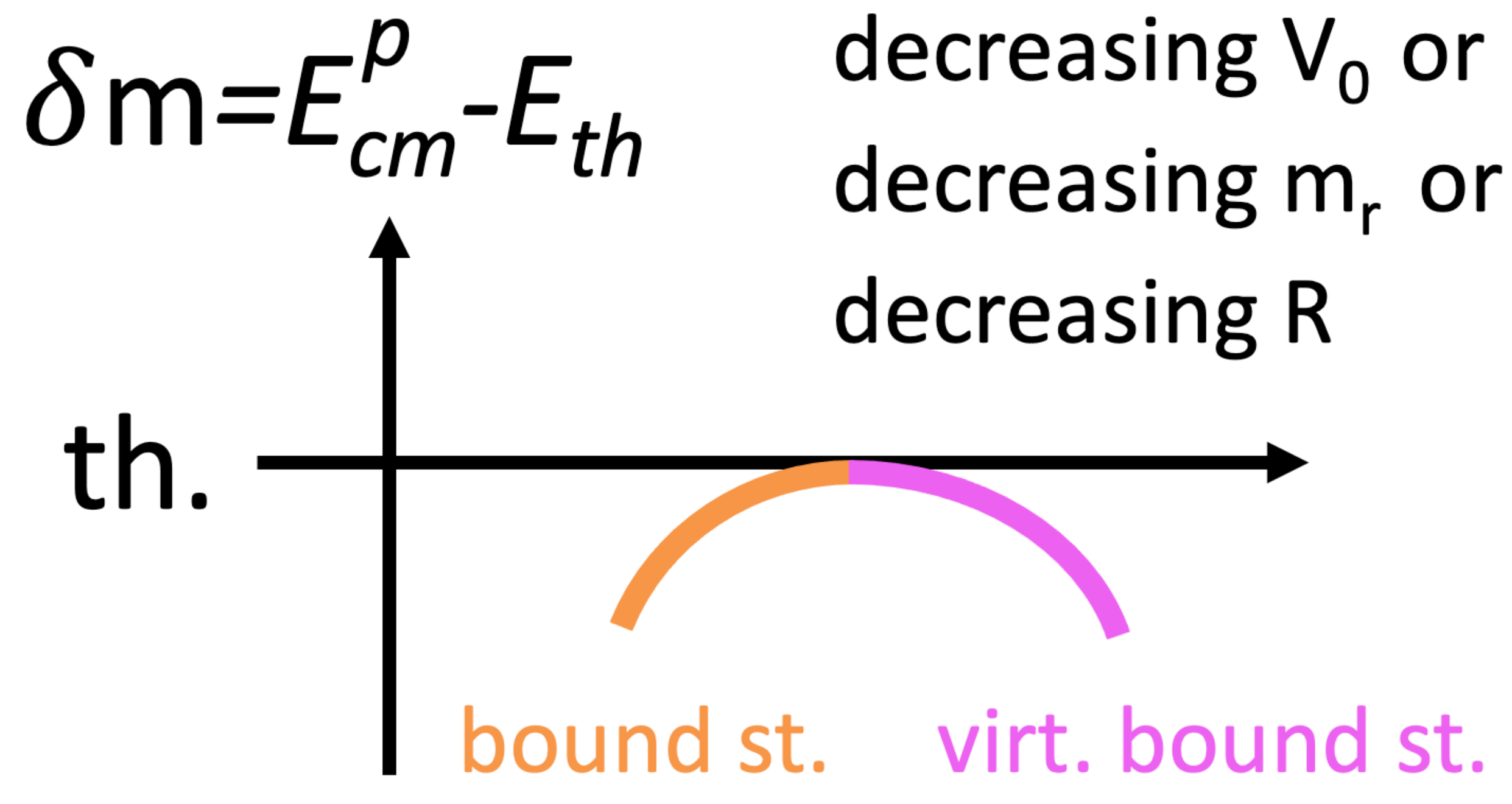}%$\qquad$\includegraphics[width=0.2\textwidth]{./mq-dep.pdf}
\caption{Sketch of the binding energy for the (virtual) bound state dominated by the molecular component 
in quantum mechanics with a purely attractive potential: $V_0$ is its overall size, $R$ is the range and 
$m_r$ is the reduced mass.}
\label{fig:mq-dep-app}
\end{center}
\end{figure}

We investigate position of the pole for purely attractive potential $V(r)=-V_0 f(r)$ between $D$ and $D^*$ 
with $V_0>0$. We explored various shapes, for example the Yukawa $f(r)=e^{-r/R}/r$, exponential $f(r)= e^{-r/R}$, 
square-well $f(r)=\theta(R-r)$. For a given potential, we numerically  determined the phase-shift $\delta_0$, 
the scattering amplitude and the energy, where scattering amplitude has a pole. We focused on the potentials 
where the pole is close to the threshold. The qualitative behavior for all potentials is analogous to the 
one derived analytically in the previous subsection for the square-well potential, as also shown in 
the video \cite{Supplv}.  A typical dependence of the pole position on varying one of the parameters in 
potential ($V_0,~R$) or the reduced mass $m_r$ is shown in Fig. \ref{fig:mq-dep-app}.  The bound state 
is present for large attraction $V_0$. As $V_0$ is decreased, the binding energy of the bound state decreases, 
at critical $V_0$ it turns to a virtual bound state and then the pole moves further below threshold. Analogous
behavior is observed when  $m_r$ or $R$  are decreased. Note that the bound state does not turn to a resonance 
for a purely attractive potential since there is no barrier to keep the resonance metastable.

Let us now consider how the quark masses affect the values of $V_0$, $R$, $m_r$ and thereby the pole positions. 
As $m_{u/d}$ increases, the mass of the exchanged light mesons $M$  also increases and the range of the potential 
$R\simeq 1/M$ decreases, which renders the dependence on $m_{u/d}$ sketched in Fig. 4 of the main article. 
Here we assumed that the dependence of the reduced mass on $m_{u/d}$ is negligible. 
   
The decrease of $m_c$ will decrease the reduced mass $m_r$ of the $DD^*$ system, while the potential will not 
change drastically due to the heavy quark flavor symmetry. The bound state becomes less and less bound with 
decreasing $m_c$ and eventually turns to a virtual bound state. 
As $m_c$ is decreased further, the virtual 
bound state pole moves further below threshold and the influence of this pole on the scattering above threshold   
is diminished.
       
       \vspace{0.1cm}

Now let us turn to quark mass dependence of the bound state dominated by the {\it diquark antidiquark Fock 
component} $[QQ][\bar u\bar d]$. The attractive colour Coulomb potential between two heavy quarks in diquark 
$[QQ]$ is flavor blind, while the kinetic energy increases with decreasing $m_Q$. This implies that 
the binding energy decreases with decreasing $m_Q$. The attraction within the good light diquark $[ud]$ 
becomes less significant as $m_{u/d}$ increases \cite{Francis:2021vrr}, which  implies that the binding energy will decrease. 
This is in line with the sketch for the bound state behavior in Fig.~4 of the main article, which is 
supported by the phenomenological studies \cite{Karliner:2017qjm,Janc:2004qn} and lattice studies of 
$bb\bar u\bar d$ \cite{Junnarkar:2018twb,Pflaumer:2021ong,Francis:2016hui,Francis:2018jyb,Leskovec:2019ioa,Frances:2022,Bicudo:2015vta}.
 However, it is not known from lattice simulations yet what is the fate of a pole when a diquark antidiquark 
state is on the verge of binding and whether it would turn to a resonance or virtual state. The analytical 
considerations of compositeness in Ref. \cite{Matuschek:2020gqe} suggest that compact states exist as bound 
states or resonances.